\documentclass[aps,amsmath,amssymb,prd,superscriptaddress,nofootinbib]{revtex4}
\usepackage{amsmath}
\usepackage{bm}
\usepackage[dvips]{graphicx}
\newcommand{\be}{\begin{equation}}

\newcommand{\ee}{\end{equation}}
\newcommand{\bea}{\begin{eqnarray}}
\newcommand{\eea}{\end{eqnarray}}

\begin{document}

\title{Electrical neutrality and pion modes in the two flavor PNJL model}
\author{H. Abuki}\email{hiroaki.abuki@ba.infn.it}
\affiliation{I.N.F.N., Sezione di Bari, I-70126 Bari, Italia}
\author{M. Ciminale}\email{marco.ciminale@ba.infn.it}
\affiliation{I.N.F.N., Sezione di Bari, I-70126 Bari, Italia}
\affiliation{Universit\`a di Bari, I-70126 Bari, Italia}
\author{R.Gatto}\email{raoul.gatto@physics.unige.ch}
\affiliation{D\'epartement de Physique Th\'eorique, Universit\'e
de Gen\`eve, CH-1211 Gen\`eve 4, Suisse}
\author{N. D. Ippolito}\email{nicola.ippolito@ba.infn.it}
\affiliation{I.N.F.N., Sezione di Bari, I-70126 Bari, Italia}
\affiliation{Universit\`a di Bari, I-70126 Bari, Italia}
\author{G. Nardulli}\email{giuseppe.nardulli@ba.infn.it}
\affiliation{I.N.F.N., Sezione di Bari, I-70126 Bari, Italia}
\affiliation{Universit\`a di Bari, I-70126 Bari, Italia}
\author{M. Ruggieri}\email{marco.ruggieri@ba.infn.it}
\affiliation{I.N.F.N., Sezione di Bari, I-70126 Bari, Italia}
\affiliation{Universit\`a di Bari, I-70126 Bari, Italia}

\begin{abstract}
We study the phase diagram and the pion modes in the electrically
neutral two flavor PNJL model. One of the main result of this
paper is that in the model with massive quarks, when electrical
neutrality is required, pions do not condense in the ground state
of the model: the isospin chemical potential $\mu_I = -\mu_e/2$ is
always smaller than the value required for pion condensation to
occur. Moreover we investigate on the pions and $\sigma$ mass
spectra. We find that the qualitative behavior of the masses
resembles that obtained in the NJL model. We close this paper by
studying the intriguing possibility that bound states with the
quantum numbers of the pions can be formed above the chiral phase
transition.
\end{abstract}

\maketitle
\section{Introduction}
The use of effective models to understand the phases of Quantum
Chromodynamics (QCD) is nowadays a popular tool. Among them, the
Nambu-Jona Lasinio model (NJL) is widely used since it allows for
a simple discussion of chiral symmetry
breaking~\cite{Nambu:1961tp,revNJL}. The main defect of the NJL
model is that it completely ignores glue dynamics: gluons are not
introduced in the lagrangian, and the QCD quark-quark interaction
mediated by gluons is replaced by an effective four fermion
interaction.

An improvement of the NJL model is provided by the introduction of a background temporal static gluon field, coupled to
quarks via the QCD covariant derivative. The gluon field is related to the expectation value of Polyakov loop
$\Phi$~\cite{Polyakovetal}, and the new model is called Polyakov-NJL (PNJL in the
following)~\cite{Meisinger:1995ih,Fukushima:2003fw,Ratti:2005jh,Roessner:2006xn,Ghosh:2007wy,
Kashiwa:2007hw,Schaefer:2007pw,Ratti:2007jf,Sasaki:2006ww,Megias:2006bn,Zhang:2006gu,Ciminale:2007ei,Fu:2007xc,Ciminale:2007sr}
It is well known that the Polyakov loop serves as an order parameter for confinement $\rightarrow$ deconfinement
transition in the pure glue theory~\cite{Polyakovetal}. In more detail, the confined phase is characterized by
$\langle\Phi\rangle=0$ and a $Z_3$ symmetry; on the other hand in the deconfined phase $\langle\Phi\rangle\neq0$ and
the symmetry $Z_3$ is broken. In presence of dynamical quarks, $Z_3$ symmetry is explicitly broken and $\Phi$ can not
be used as an order parameter. Nevertheless it is commonly used as an indicator of the deconfinement transition.

In the PNJL model an effective potential for $\Phi$ is added by
hand to the quark lagrangian, and $\Phi$ is coupled to the quarks
via the QCD covariant derivative. The value of
$\langle\Phi\rangle$ in the ground state, as well as other
quantities of interest like the constituent quark masses, are
obtained by minimization of the thermodynamic potential, or
equivalently by solving coupled gap equations as in the NJL case.
Despite the apparent complication due to the increase of the
degrees of freedom, the PNJL model offers a better description of
QCD then NJL does since it allows to derive, in the framework of
field theory, several results obtained in lattice
simulations~\cite{Ratti:2005jh,Roessner:2006xn,Ghosh:2007wy,
Kashiwa:2007hw,Ratti:2007jf,Sasaki:2006ww,Zhang:2006gu}.

The knowledge of the phase diagram of the PNJL model as well as of its bulk properties is of a certain interest. In
this paper we study the phase diagram and the pion modes in the electrically neutral two flavor PNJL model. The
motivation of our study is straightforward: we are interested to investigate the equilibrium ground state of the model,
and in the ground state a net electric charge can not be present, whichever the value of the temperature and of the
quark chemical potential is.   We consider the possibility of chiral as well as pion condensation in the ground state,
at zero and non zero quark chemical potential. A related study without electrical neutrality and at zero chemical
potential has been done in Ref.~\cite{Zhang:2006gu}; see also Ref.~\cite{Ebert:2005wr} for a similar study in the
framework of the electrically neutral NJL model in the chiral limit.

One of the results of our work is that when electrical neutrality is required, pion do not condense in the ground state
of the model: the isospin chemical potential $\mu_I = -\mu_e/2$, with $\mu_e$ the electron chemical potential, is
always smaller than the value required for pion condensation to occur. Moreover we investigate on the pions and
$\sigma$ mass spectra. We find that the qualitative behavior of the masses resembles that obtained in the NJL model. We
close this paper by studying the possibility that a bound state with the quantum numbers of the pions can be formed
above the chiral phase transition.

The plan of the paper is as follows: in Section~\ref{Sec:for} we
briefly describe the PNJL model; in Section~\ref{Sec:pd} we show
the phase diagram of the model, and some results for quantities of
interest (in medium quark mass, pion condensate, mean value of the
Polyakov loop, electron chemical potential). In
Section~\ref{Sec:PGm} we compute the masses of the
pseudo-Goldstone modes of the model and of the $\sigma$ mode.
Finally in Section~\ref{Sec:con} we draw our conclusions.

\section{Formalism\label{Sec:for}}
The Lagrangian density of the two flavor PNJL model is given by~\cite{Fukushima:2003fw}
\begin{equation}
{\cal L}^\prime= \bar{e}(i\gamma_\mu\partial^\mu)e +  \bar\psi\left(i\gamma_\mu D^\mu + \mu\gamma_0 -m\right)\psi +
G\left[\left(\bar\psi \psi\right)^2 + \left(\bar\psi i \gamma_5 \vec\tau \psi\right)^2\right] - {\cal
U}[\Phi,\bar\Phi,T]~. \label{eq:LagrP}
\end{equation}
In the above equation $e$ denotes the electron field;  $\psi$ is the quark spinor with Dirac, color and flavor indices
(implicitly summed). $m$ corresponds to the bare quark mass matrix; we assume from the very beginning $m_u = m_d$. The
covariant derivative is defined as usual as $D_\mu =
\partial_\mu -i A_\mu$. The gluon background field
$A_\mu=\delta_{0\mu}A_0$ is supposed to be homogeneous and static,
with $A_0 = g A_0^a T_a$ and $T_a$, $a=1,\dots,8$ being the
$SU(3)$ color generators with the normalization condition
$\text{Tr}[T_a T_b]=\delta_{ab}$. $\vec{\tau}$ is a vector of
Pauli matrices in flavor space. Finally $\mu$ is the chemical mean
quark chemical potential, related to the conserved baryon number.

In Eq.~\eqref{eq:LagrP} $\Phi$, $\bar\Phi$ correspond to the normalized traced Polyakov loop $L$ and its hermitian
conjugate respectively, $\Phi=\text{Tr}L/N_c$, $\bar\Phi=\text{Tr}L^\dagger/N_c$, with
\begin{equation}
L={\cal P}\exp\left(i\int_0^\beta A_4 d\tau\right)=\exp\left(i \beta A_4\right)~,~~~~~A_4=iA_0~,
\end{equation}
and $\beta=1/T$. The term ${\cal U}[\Phi,\bar\Phi,T]$ is the effective potential for the traced Polyakov loop; in
absence of dynamical quarks it is built in order to reproduce the pure glue lattice data of QCD, namely thermodynamical
quantities (pressure, entropy and energy density) and the deconfinement temperature of heavy (non-dynamical) quarks,
$T= 270$ MeV. Several forms of this potential have been suggested in the literature, see for
example~\cite{Fukushima:2003fw,Ratti:2005jh,Roessner:2006xn,Ghosh:2007wy}. In this paper we adopt the following
logarithmic form~\cite{Roessner:2006xn},
\begin{equation}
\frac{{\cal U}[\Phi,\bar\Phi,T]}{T^4} = -\frac{b_2(T)}{2}\bar\Phi\Phi + b(T)\log\left[1-6\bar\Phi\Phi + 4(\bar\Phi^3 +
\Phi^3) -3(\bar\Phi\Phi)^2\right]~,\label{eq:Poly}
\end{equation}
with
\begin{equation}
b_2(T) = a_0 + a_1 \left(\frac{\bar T_0}{T}\right) + a_2 \left(\frac{\bar T_0}{T}\right)^2~,~~~~~b(T) =
b_3\left(\frac{\bar T_0}{T}\right)^3~.\label{eq:lp}
\end{equation}
Numerical values of the coefficients are as follows~\cite{Roessner:2006xn}:
\begin{equation}
a_0=3.51~,~~~a_1 = -2.47~,~~~a_2 = 15.2~,~~~b_3=-1.75~.
\end{equation}
If dynamical quarks were not present then one should chose $\bar T_0 = 270$ MeV in order to reproduce the deconfinement
transition at $T = 270$ MeV.  In presence of quarks $\bar T_0$ gets a dependence on the number of active flavors, as
shown in Refs.~\cite{Ratti:2005jh,Schaefer:2007pw}. Following Ref.~\cite{Schaefer:2007pw} we chose $\bar T_0 = 208$ MeV
in Eq.~\eqref{eq:lp}, which is appropriate to deal with two degenerate flavors.

As explained in the Introduction we are interested to study the ground state of the model specified by the Lagrangian
in Eq.~\eqref{eq:LagrP}, at each value of the temperature $T$ and the chemical potential $\mu$,  corresponding to a
vanishing total electric charge. To this end we introduce electrons in Eq.~\eqref{eq:LagrP} since a net number of
electrons could be needed, beside the pion condensate (if any, see below), to ensure the electrical neutrality of the
ground state. In order to build the neutral ground state we work as usual in the gran canonical ensemble formalism,
adding to Eq.~\eqref{eq:LagrP} the term $\mu_Q N_Q$, $\mu_Q$ being the chemical potential  for the total charge $N_Q$,
and requiring stationarity of the thermodynamic potential with respect to variations of $\mu_Q$, which is equivalent to
the requirement $<N_Q>=0$ in the ground state. To be more specific, since the total charge operator is given in terms
of the quark and electron fields by
\begin{equation}
N_Q = \frac{2}{3}N_u - \frac{1}{3}N_d - N_e = \frac{2}{3}{u^\dagger}u - \frac{1}{3}{d^\dagger}d -{e^\dagger}e~,
\end{equation}
it is easy to recognize that the lagrangian ${\cal L}$ in the gran canonical ensemble ${\cal L} = {\cal L}^\prime +
\mu_Q N_Q$ can be written as
\begin{equation}
{\cal L}=\bar{e}(i\gamma_\mu\partial^\mu + \mu_e \gamma_0)e +  \bar\psi\left(i\gamma_\mu D^\mu + \hat\mu\gamma_0
-m\right)\psi + G\left[\left(\bar\psi \psi\right)^2 + \left(\bar\psi i \gamma_5 \vec\tau \psi\right)^2\right] - {\cal
U}[\Phi,\bar\Phi,T]~, \label{eq:Lagr}
\end{equation}
where $\mu_e = - \mu_Q$ and the quark chemical potential matrix $\hat\mu$ is defined in flavor-color space as
\begin{equation}
\hat\mu=\left(\begin{array}{cc}
  \mu-\frac{2}{3}\mu_e & 0 \\
  0 & \mu + \frac{1}{3}\mu_e \\
\end{array}\right)\otimes\bm{1}_c~,\label{eq:chemPot}
\end{equation}
where $\bm{1}_c$ denotes identity matrix in color space.

In this paper we work in the mean field approximation. In order to study chiral symmetry breaking and to allow for pion
condensation we assume that in the ground state the expectation values for the following operators may
develop~\cite{Zhang:2006gu,Ebert:2005wr},
\begin{equation}
\sigma = \left<\bar\psi \psi\right>~,~~~~~\pi = \left<\bar\psi i \gamma_5 \tau_1 \psi\right>~.\label{eq:condensates}
\end{equation}
In the above equation a summation over flavor and color is understood.  We have assumed that the pion condensate aligns
along the $\tau_1$ direction in flavor space. This choice is not restrictive. As a matter of fact we should allow for
independent condensation both in $\pi^+$ and in $\pi^-$ channels~\cite{Zhang:2006gu}:
\begin{equation}
\pi^+\equiv\langle\bar\psi i \gamma_5 \tau_+ \psi\rangle =
\frac{\pi}{\sqrt{2}}e^{i\theta}~,~~~~~\pi^-\equiv\langle\bar\psi i \gamma_5 \tau_- \psi\rangle=
\frac{\pi}{\sqrt{2}}e^{-i\theta}~,
\end{equation}
with $\tau_\pm = (\tau_1\pm\tau_2)/\sqrt{2}$; but the thermodynamical potential potential is not dependent on the phase
$\theta$, therefore we can assume $\theta=0$ which leaves us with $\pi^+ = \pi^- = \pi/\sqrt{2}$ and introduce only one
condensate, specified in Eq.~\eqref{eq:condensates}. In the mean field approximation the PNJL lagrangian reads
\begin{equation}
{\cal L}=\bar{e}(i\gamma_\mu\partial^\mu + \mu_e \gamma_0)e + \bar\psi\left(i\gamma_\mu D^\mu + \hat\mu\gamma_0 -M - N
i\gamma_5\tau_1\right)\psi - G\left[\sigma^2 + \pi^2\right] - {\cal U}[\Phi,\bar\Phi,T]~, \label{eq:LagrMF}
\end{equation}
where the in-medium quark mass $M$ and the pion field $N$ are
related to $\sigma$ and $\pi$ by means of the relations
\begin{eqnarray}
M&=&m-2G\sigma~,\\
N&=&-2G\pi~.
\end{eqnarray}
The thermodynamical potential $\Omega$ can be obtained by integration over the fermion fields in the partition function
of the model, see for example Ref.~\cite{Ebert:2000pb},
\begin{equation}
\Omega = -\left(\frac{\mu_e^4}{12\pi^2} + \frac{\mu_e^2 T^2}{6} + \frac{7\pi^2 T^4}{180}\right) + {\cal
U}[\Phi,\bar\Phi,T] + G\left[\sigma^2 + \pi^2\right] - T\sum_n\int_0^\Lambda\frac{d^3{\bm
p}}{(2\pi)^3}~\text{Tr}~\text{log}\frac{S^{-1}(i\omega_n,{\bm p})}{T}~,
\end{equation}
where the sum is over fermion Matsubara frequencies $\omega_n = \pi T(2n+1)$, and the trace is over Dirac, flavor and
color indices. $\Lambda$ is an ultraviolet cutoff to ensure convergence of the momentum integral. The inverse quark
propagator is defined as
\begin{equation}
S^{-1}(i\omega_n,{\bm p})= \left(\begin{array}{cc}
  (i\omega_n+\mu-\frac{2}{3}\mu_e+iA_4)\gamma_0 -{\bm\gamma}\cdot{\bm p} -M & -i\gamma_5 N \\
  -i\gamma_5 N & (i\omega_n+\mu+\frac{1}{3}\mu_e+iA_4)\gamma_0 -{\bm\gamma}\cdot{\bm p} -M\\
\end{array}\right)\otimes{\bm 1}_c~.\label{eq:po}
\end{equation}
The ground state of the model is defined by the values of
$\sigma$, $\pi$, $\Phi$, $\bar\Phi$ that minimize $\Omega$ and
that have a vanishing total charge; the latter condition is
equivalent to the requirement
\begin{equation}
\frac{\partial\Omega}{\partial\mu_e}=0~.
\end{equation}
The parameters $m$, $G$ and $\Lambda$ are given by~\cite{Roessner:2006xn}
\begin{equation}
m=5.5~\text{MeV}~,~~~~~G=5.04~\text{GeV}^{-2}~,~~~~~\Lambda=650.9~\text{MeV}~,
\end{equation}
which fix, at zero temperature and zero chemical potential, the
pion mass $m_\pi=139.3$ MeV, the pion decay constant $f_\pi=92.3$
MeV and the chiral condensate $\langle\bar u u\rangle =
-(251~\text{MeV})^3$.

\section{Phase diagram of the model\label{Sec:pd}}
\begin{figure}
\begin{center}
\includegraphics[width=8cm]{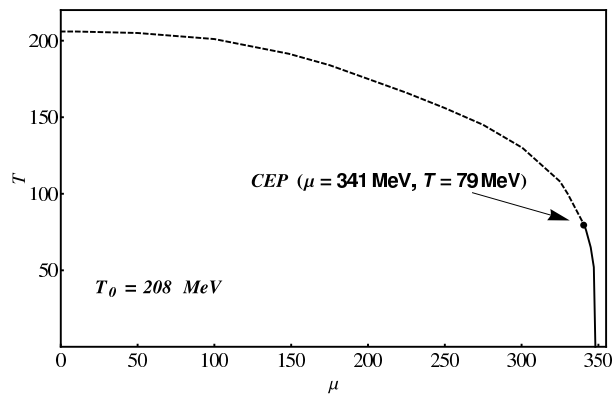}
\end{center}
\caption{\label{FIG:pd}Phase diagram of the electrically neutral
two flavor PNJL model. Dashed line corresponds to the chiral
crossover; solid line describes the first order chiral transition.
At $\mu=0$ the critical point is located at $T=206$ MeV; at $T=0$
the chiral transition is found at $\mu=348$ MeV. The black dot
denotes the critical end point (CEP), located at $(\mu_E,T_E) =
(342,79)$ MeV.}
\end{figure}

In Fig.~\ref{FIG:pd} we plot the phase diagram of the electrically
neutral two flavor PNJL model. Dashed line corresponds to the
chiral crossover; solid line describes the first order chiral
transition. At each value of $\mu$ the crossover is identified
with the inflection point of the chiral condensate. Analogously
the first order transition is defined by the discontinuity of
$\sigma$. In the region below the lines the chiral symmetry is
broken, above the lines it is restored. At $\mu=0$ the chiral
symmetry is restored at $T=206$ MeV. For comparison, the
inflection point of the Polyakov loop (which is commonly
associated to the deconfinement crossover) at $\mu=0$ is located
at $T=180$ MeV. Moreover, at $T=0$ the chiral phase transition is
of first order and is found at $\mu=348$ MeV. The black dot
denotes the critical end point (CEP), located at $(\mu_E,T_E) =
(342,79)$ MeV. As explained later, we find a vanishing pion
condensate once electrical neutrality is required.

\begin{figure}
\begin{center}
\includegraphics[width=10cm]{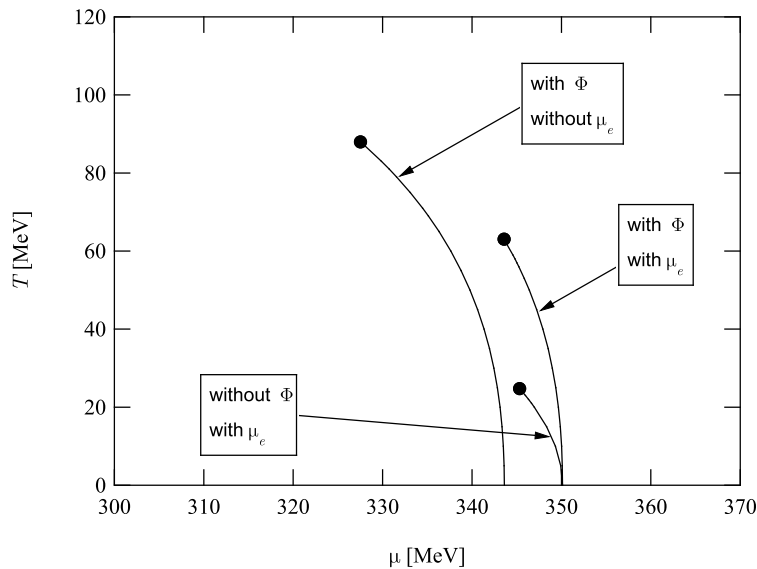}
\end{center}
\caption{\label{FIG:pd2}Comparison of the first order transitions
of the PNJL model with neutrality (called Model I in the text), of
the PNJL model with $\mu_e=0$ (model II) and of the NJL model with
electrical neutrality (model III).}
\end{figure}

It is instructive to compare the PNJL results discussed above with
the electrically neutral NJL model, as well as the PNJL model with
$\mu_e=0$, in order to emphasize the role of electrical neutrality
and of the Polyakov loop. To this end in Fig.~\ref{FIG:pd2} we
compare the first order transitions of the PNJL model with
neutrality (model I), of the PNJL model with $\mu_e=0$ (model II)
and of the NJL model with electrical neutrality (model III).
Comparison of models I and II shows that adding the electron
chemical potential and requiring electrical neutrality enlarges of
some MeV the $\mu$ window of the chirally broken phase. Moreover,
comparison of models I and III shows that adding the
self-consistently evaluated Polyakov loop to the model renders the
broken phase more robust, as the critical temperatures are
increased.

\begin{figure}
\begin{center}
\includegraphics[width=7cm]{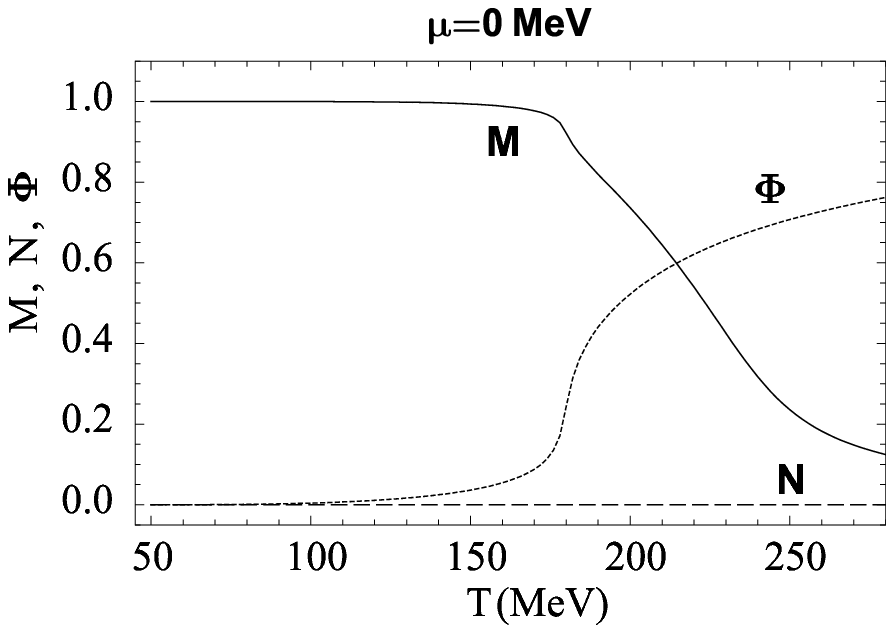}~~~~~\includegraphics[width=7cm]{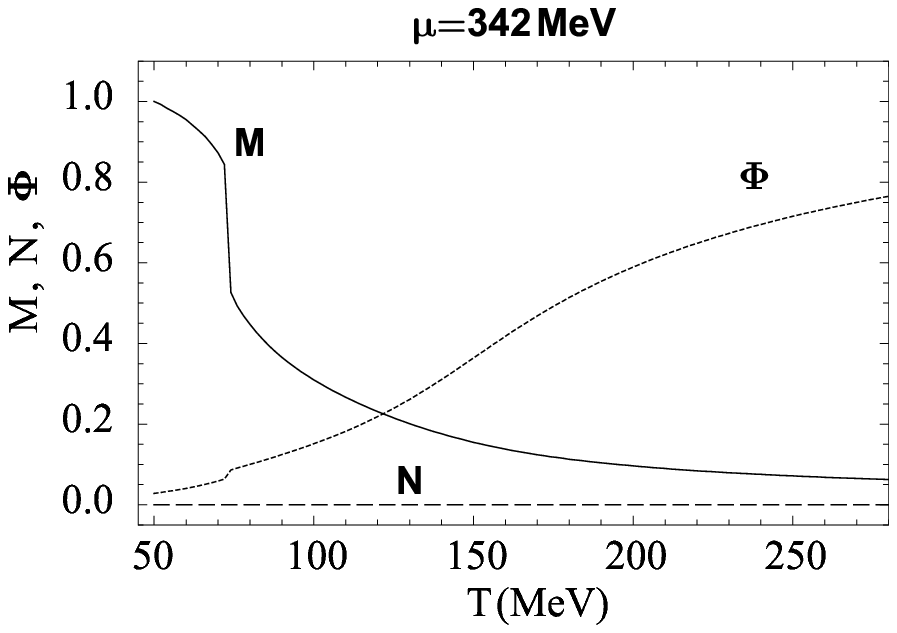}
\end{center}
\caption{\label{FIG:nn}Quark constituent mass $M$ (in units of
$M_0$), pion condensate $N$ (in units of $M_0$) and $\Phi$ as a
function of temperature (in MeV) for different values of $\mu$.
$M_0 = 325$ MeV is the constituent quark mass evaluated at
$\mu=0$, $T=0$ and $\mu_e=0$.}
\end{figure}

\begin{figure}
\begin{center}
\includegraphics[width=8cm]{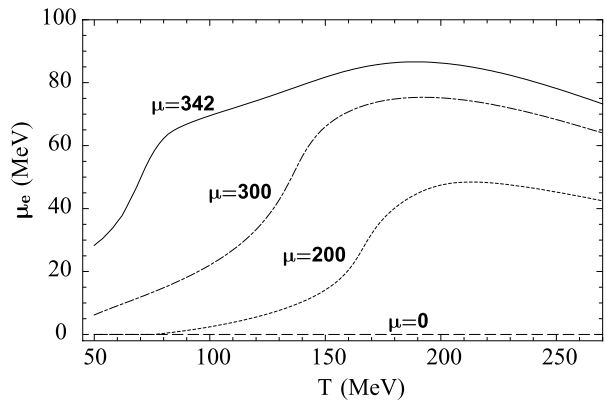}
\end{center}
\caption{\label{FIG:mue}Electron chemical potential as a function
of the temperature, for different values of the quark chemical
potential. }
\end{figure}

In Fig.~\ref{FIG:nn} we plot the quark constituent mass $M$ (in units of $M_0$), pion condensate $N$ (in units of
$M_0$) and $\Phi$ as a function of temperature (in MeV) for two values of $\mu$. We have defined $M_0 = 325$ MeV as the
constituent quark mass evaluated at $\mu=0$, $T=0$ and $\mu_e=0$. The electron chemical potential evaluated at some
values of $\mu$ is shown in Fig.~\ref{FIG:mue}. We observe that when electrical neutrality is required, pions do not
condense in the ground state of the model. As a matter of fact $N=0$ in Fig.~\ref{FIG:nn}; we have verified this result
for several values of $\mu$. Stated in other words the isospin chemical potential $\mu_I = -\mu_e/2$ in the neutral
phase is always smaller than the value required for pion condensation to occur.

\begin{figure}
\begin{center}
\includegraphics[width=8cm]{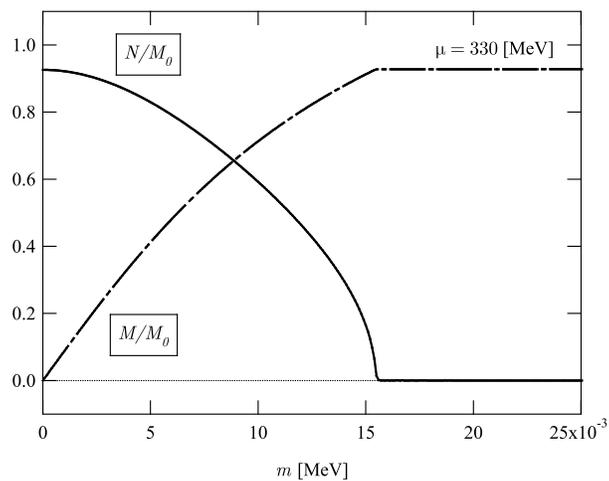}
\end{center}
\caption{\label{FIG:cmd}Pion condensate $N$ and the constituent quark mass $M$ in units of $M_0$, the constituent quark
mass at $T=\mu=0$, as a function of the bare quark mass $m_0$ at the representative value of $\mu=330$ MeV and $T=0$.}
\end{figure}

We have verified the robustness of our calculations in several
ways. First, we have treated $\mu_e$ as a free parameter,
reproducing the results of Ref.~\cite{Zhang:2006gu} finding a pion
condensate for high values of $\mu_e$. Second, we have compared
our results with those Ref.~\cite{Ebert:2005wr}. The authors of
Ref.~\cite{Ebert:2005wr} study the neutral ground state in the NJL
model at low temperature. They use a set of parameters in which
the values of $G$ and $\Lambda$ are similar to ours, but they
consider only the chiral limit $m_u = m_d = 0$. In this limit they
find a narrow window in $\mu$ at zero temperature in which the
pion condensate can exist. We have explicitly checked that putting
by hand $m_u = m_d = 0$ in our calculations we reproduce the
aforementioned window. This is a further check since at low
temperature the PNJL reduces to the NJL model. Moreover we have
verified that the pion condensate is washed out if the current
quark mass is increased from zero to the physical value $m_u = m_d
= 5.5$ MeV. This is explicitly shown in Fig~\ref{FIG:cmd} where we
plot the pion condensate $N$ and the constituent quark mass $M$ in
units of $M_0$, the constituent quark mass at $T=\mu=0$, as a
function of the bare quark mass $m_0$ at the representative value
of $\mu=330$ MeV (we have checked that the same result hold for
other values of $\mu$) and $T=0$. Therefore the current quark mass
works as an external field that drives the alignment of the vacuum
along the chiral condensate direction. We stress that Fig.~5 has
to be taken only as a trick that allows a clear comparison between
our results and those of Ref.~\cite{Ebert:2005wr}. A true plot of
$M$ and $N$ as a function of the current quark mass requires the
self consistent evaluation of $G$ and $\Lambda$, which is beyond
the scope of the present paper since we study only the phase
diagram and the pion modes at the physical point. Nevertheless the
problem of the modification of the ground state by variations of
the current quark mass is interesting and deserves further study,
therefore we leave this to a future project.

\section{Pseudo-Goldstone modes\label{Sec:PGm}}
In this Section we compute the masses of the pseudo-Goldstone
modes as well as of the $\sigma$ mode in the two flavor and
electrically neutral PNJL model. As discussed in the previous
Section for each value of $\mu$ and $T$, after requiring
electrical neutrality, the difference of chemical potentials
between up and down quarks is lower than the critical value
required for the realization of the pion condensation. In this
Section we therefore put by hand $N=0$ in the quark propagator,
Eq.~\eqref{eq:po}.

The pion mass is defined as the solution of the well-known
Bethe-Salpeter equation for the bound states. Its derivation in
the context of the NJL model is in Ref.~\cite{revNJL}. The same
equation is valid in the PNJL model, the difference between PNJL
and NJL being only the different quark propagator to be used in
the calculation of the pion polarization tensor. The equation
reads
\begin{equation}
1-2G\Pi_A(m^2_A)=0~,~~~~~A=1,2,3~.\label{eq:bs1}
\end{equation}
In the above equation $\Pi_A$ is the polarization tensor of the
pion $A$, specified later, where $A=1,2,3$ correspond respectively
to $\pi^+$, $\pi^-$ and $\pi^0$; $m_A$ denotes the pion mass, and
$G$ is the coupling constant introduced in Eq.~\eqref{eq:Lagr}.
The solutions of Eq.~\eqref{eq:bs1} correspond to the poles of the
pion propagator, the latter evaluated in the random phase
approximation~\cite{revNJL}.

We now specify the pion polarization tensor. We introduce the
polarization matrix
\begin{equation}
\Pi_{ij}(m^2) = -i\int\frac{d^4p}{(2\pi)^4}\text{Tr}\left[i\gamma_5 \tilde\tau_i iS\left(p+\frac{k}{2}\right)i\gamma_5
\tilde\tau_j iS\left(p-\frac{k}{2}\right)\right]~;\label{eq:pp}
\end{equation}
here $i,j=1,2,3$, $k_\mu = (m,{\bm 0})$, and the trace is over color, flavor and Dirac indices; the matrices
$\tilde\tau_i$ are operators in the flavor space, defined by
\begin{equation}
\tilde\tau_{1,2}=\frac{1}{\sqrt{2}}\left(\tau_1\pm i \tau_2\right)~,~~~~~ \tilde\tau_3 = \tau_3~,
\end{equation}
where $\tilde\tau_i$, $i=1,2,3$ are the Pauli matrices; $S$ is the fermion propagator, whose inverse is defined in
Eq.~\eqref{eq:po} with $N=0$, and masses and chemical potentials evaluated self consistently in the electrically
neutral phase. In terms of the matrix $\Pi_{ij}$ the pion polarization tensors appearing in Eq.~\eqref{eq:pp} are
defined as
\begin{equation}\Pi_1 = \Pi_{12}~,~~~~~
\Pi_2= \Pi_{21}~,~~~~~\Pi_3 = \Pi_{33}~.
\end{equation}

Analogously the equation for the $\sigma$ mass reads
\begin{equation}
1-2G\Pi_\sigma(m^2_\sigma)=0~,\label{eq:bss}
\end{equation}
with the polarization tensor defined as
\begin{equation}
\Pi_\sigma(m^2) = -i\int\frac{d^4p}{(2\pi)^4}\text{Tr}\left[
iS\left(p+\frac{k}{2}\right)
iS\left(p-\frac{k}{2}\right)\right]~.\label{eq:pps}
\end{equation}

Once the traces in Eqs.~\eqref{eq:bs1} and~\eqref{eq:bss} are
evaluated, we are left with the loop 4-momentum integral. We
replace the integral over energy by a summation over fermion
Matsubara frequencies; moreover, the integral over directions of
momentum is trivial. Therefore we are left with the integration
over the modulus of ${\bm p}$ only, which is evaluated
numerically.

\begin{figure}
\begin{center}
\includegraphics[width=7cm]{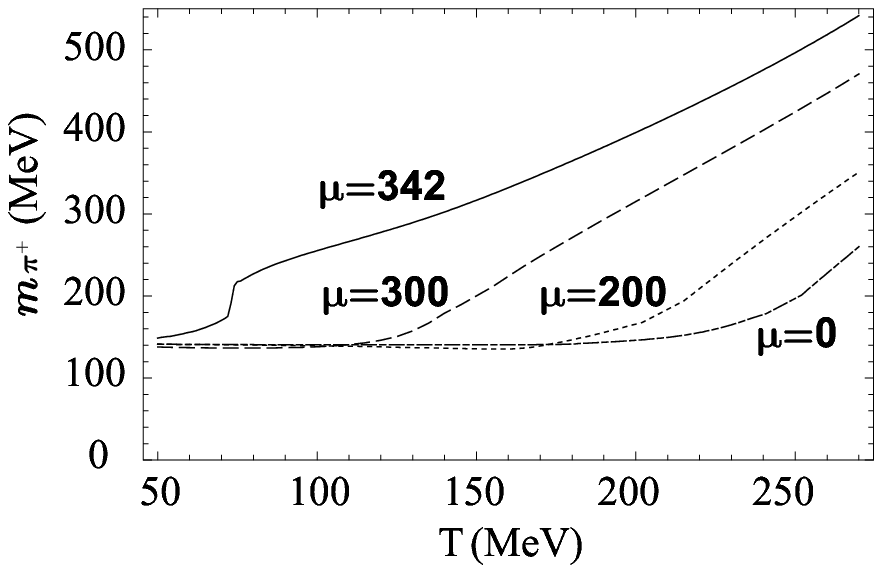}
~~~~~\includegraphics[width=7cm]{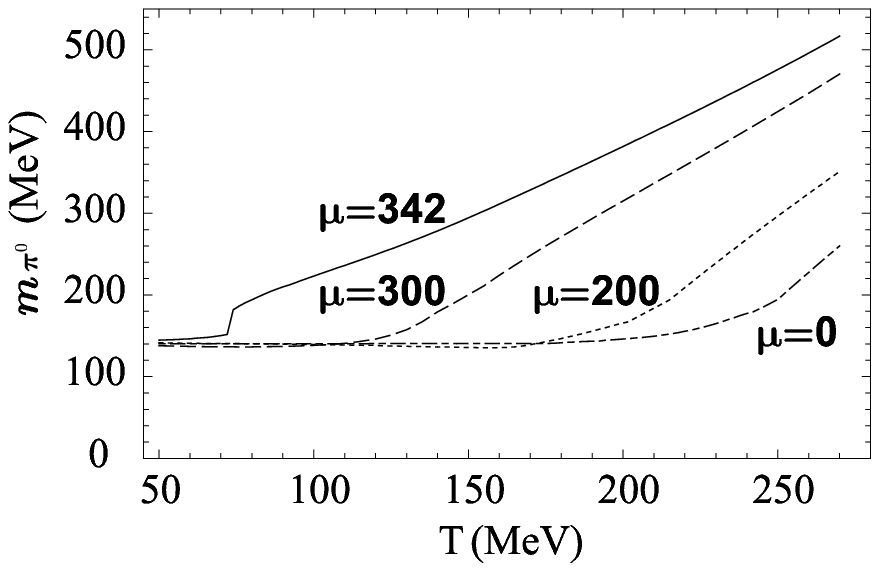}
\end{center}
\caption{\label{FIG:pp}Masses of the charged pions (left panel)
and of the neutral pion (right panel) in the electrically neutral
phase, as a function of the temperature, for different values of
the quark chemical potential.}
\end{figure}

In Fig.~\ref{FIG:pp} we plot the masses of the charged pions (left
panel) and of the neutral pion (right panel) in the electrically
neutral phase, as a function of the temperature, for different
values of the quark chemical potential. The behavior of the
pseudoscalar modes as the temperature is increased is
qualitatively the same observed in the NJL model~\cite{revNJL}. At
each value of $\mu$, the charged and neutral pion masses are of
the order of the zero temperature value $\approx 140$ MeV below
the chiral crossover (or the chiral first order transition); as
the chiral transition (either first order or crossover) occurs,
the pion masses rapidly increase signaling the disappearing of the
pseudoscalar modes from the low energy spectrum of the model.

\begin{figure}
\begin{center}
\includegraphics[width=7cm]{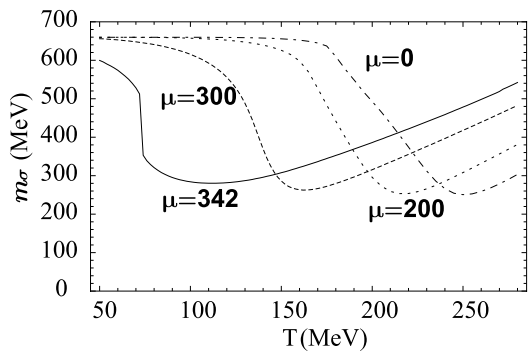}
\end{center}
\caption{\label{FIG:sm}Mass of the $\sigma$ mode in the
electrically neutral phase, as a function of the temperature, for
different values of the quark chemical potential.}
\end{figure}

In Fig.~\ref{FIG:sm} we plot the mass of the $\sigma$ mode in the
electrically neutral phase, as a function of the temperature, for
different values of the quark chemical potential. As in the case
of the pions, the behavior of $m_\sigma$ is analogous to that
observed in the NJL model~\cite{revNJL}.

\begin{figure}[h!]
\begin{center}
\includegraphics[width=7cm]{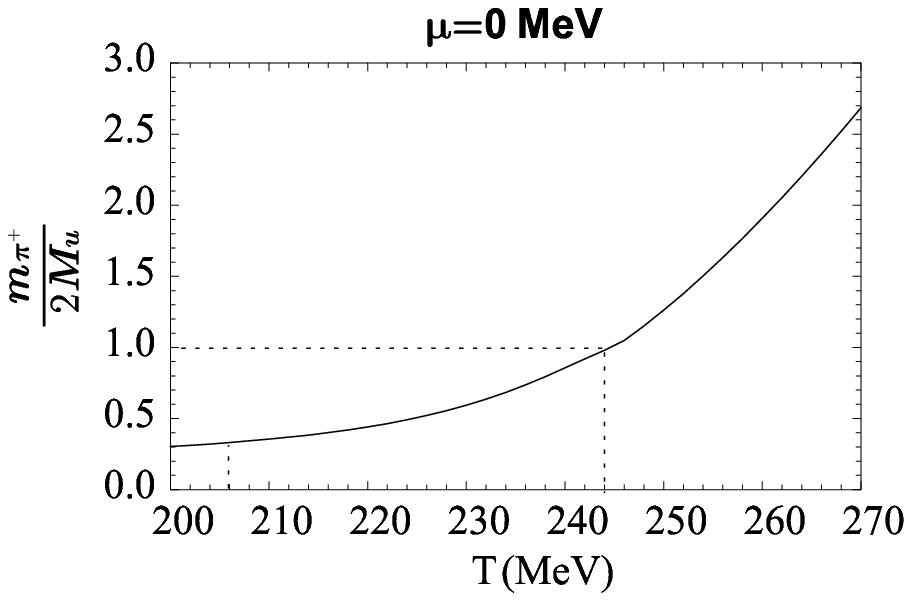}~~~~~\includegraphics[width=7cm]{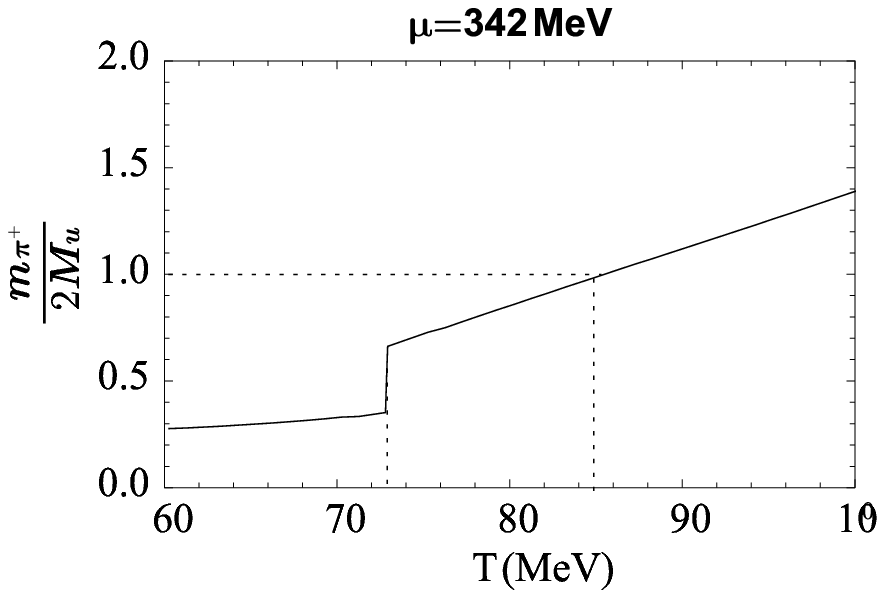}
\end{center}
\caption{\label{FIG:kkk}The ratio $m_\pi/2M$, with $M$ the
constituent quark mass, as a function of temperature, for two
representative values of the quark chemical potential. The
vertical bold dashed line denotes the chiral transition
temperature.}
\end{figure}

It is interesting to evaluate the ratio $m_\pi/2M$, with $M$ the
constituent quark mass, as a function of temperature. If it is
larger than one, than a bound state with the quantum numbers of
the pion is less stable than a state made of a free quark and a
free antiquark, and thus the pion melts to its constituent quarks.
In Fig.~\ref{FIG:kkk} we show the results of the computation of
$m_\pi/2M$ at different values of $\mu$ in the electrically
neutral phase, for values of temperatures close either to the
chiral crossover or to the first order chiral transition. We
notice that for any value of the quark chemical potential there
exists a window in temperature, above the chiral restoration
temperature, in which the pions are lighter than a quark and an
antiquark. The window shrinks as $\mu$ is increased. These results
show that even above the critical temperature a bound state with
the pions quantum numbers can be formed.

We observe that the window in which the bound state is stable above the chiral phase transition (as well as the chiral
crossover) is not a peculiarity of the PNJL model. As a matter of fact the same window was noticed in the NJL model at
$\mu=0$ in Ref.~\cite{Hatsuda:1985eb}. We have verified that it exists in the neutral NJL model as well. For example,
at $\mu=0$ we find that the bound states exist in the window $(195,212)$ MeV; this result has to be compared with the
PNJL one, namely $(205,245)$ MeV, showing that the PNJL window is almost 2.4 times larger than the NJL one. Analogously
at $\mu=342$ MeV the NJL window is $(34,44)$ MeV, to be compared with the PNJL result $(73,85)$ MeV.

Similar results have been obtained in the framework of the PNJL model in Ref.~\cite{Hansen:2006ee}. The authors of
Ref.~\cite{Hansen:2006ee} consider the two flavor PNJL model without the constraint of electrical neutrality. Moreover
they use a polynomial form of the Polyakov loop effective potential ${\cal U}$ in Eq.~\eqref{eq:Lagr} instead of the
logarithmic one, see Eq.~\eqref{eq:Poly}.  Therefore a quantitative comparison with their results is not feasible.
Instead we have computed the ratio $T_M/T_\chi$ in the case $\mu_e=0$ in the PNJL model with the logarithmic form of
the effective potential,  for several values of $\mu$, in order to quantify the role of the electrical neutrality on
the ratio $T_M/T_\chi$. Our results are summarized in Table~\ref{tab:comp}.

\begin{table}
\begin{center}
\begin{tabular}{|c|c|c|c|c|c|c|}
\hline
 &$T_\chi~ (\mu_e = 0)$& $T_M~(\mu_e = 0)$ & $T_M/T_\chi~(\mu_e=0)$ &$T_\chi$ & $T_M$ &$T_M/T_\chi$\\
 \hline
$\mu=100$ &  $218$& $236$ &1.08 &$198$ & $234$  &1.18\\
\hline $\mu=200$ &  $178$& $207$ & 1.16&$175$ & $211$  &1.20\\
\hline
$\mu=300$ &  $131$& $145$ & 1.10&$131$ & $154$  &1.17\\
\hline
$\mu=342$ &  $44$& $48$ &1.09 &$73$ & $85$  & 1.16\\
\hline
\end{tabular}
\end{center}
\caption{\label{tab:comp}Comparison of the pion melting temperatures $T_M$ and the chiral transition temperatures
$T_\chi$ with electrical neutrality and with $\mu_e = 0$. Quark chemical potentials and temperatures are measured in
MeV.}
\end{table}

\begin{figure}[h!]
\begin{center}
\includegraphics[width=10cm]{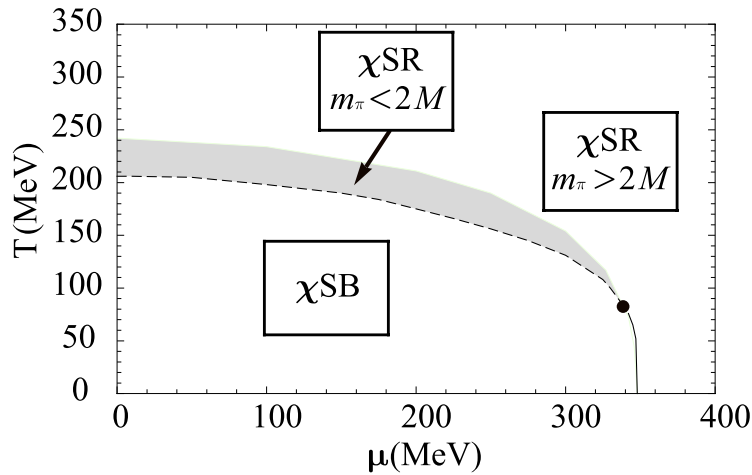}
\end{center}
\caption{\label{FIG:ccc}Region of existence of bound states in the
$\mu-T$ plane. $\chi SB$ denotes the region where chiral symmetry
is broken; $\chi SR$ denotes the region where chiral symmetry is
restored. $M$ is the constituent quark mass. The dashed and the
solid lines correspond respectively to the chiral crossover and to
the chiral first order transition, the dot denotes the CEP. In the
gray region chiral symmetry is restored but a bound state with the
quantum numbers of the pions can still be formed. See the text for
more details.}
\end{figure}

It is interesting to superimpose the bound state existence region
to the phase diagram in Fig.~\ref{FIG:pd}. This is done in
Fig.~\ref{FIG:ccc} where we show the region in the $\mu-T$ plane
 where the bound state can be formed. In the white region below the
gray domain the chiral symmetry is broken and pions live as
pseudo-Goldstone modes. In the gray region chiral symmetry is
restored but $m_\pi/2M_q < 1$, thus bound states can be formed.
Finally, in the white region above the gray domain chiral symmetry
is restored and free quark states are more stable than bound
states (this region is characterized by $m_\pi/2M_q > 1$).

The existence of bound states above the chiral transition temperature is interesting because it implies that quark
matter above the chiral transition is strongly coupled. It has been suggested~\cite{Shuryak:2003ty,Brown:2003km} that
the formation of bound states just above the chiral transition can explain the experimental results obtained by
non-central heavy ion collisions at RHIC facility~\cite{Kolb:2003dz,Molnar:2001ux,Teaney:2003kp,Nakamura:2004sy}; in
particular it provides a simple explanation to the low viscosity-to-entropy ratio observed in elliptic flow
simulations. The next step would be the computation of the transport coefficients of quark matter just above the chiral
transition temperature. We leave this to a future project.

\section{Conclusions\label{Sec:con}}
In this paper we have studied the two flavor PNJL model in
presence of a charge chemical potential, requiring electrical
neutrality in the mean field approximation. We have assumed that
in the ground state both the chiral and the pion condensates
develop, and we have evaluated their values, as well as the value
of the Polyakov loop, for each value of the temperature and of the
quark chemical potential by minimizing the thermodynamic potential
under the condition that the total electric charge of the system
vanishes.

We have drawn the phase diagram of the model, see
Fig.~\ref{FIG:pd}. One of our main results is that we do not
observe pion condensation in the mean field approximation. We
relate this to the non vanishing current quark mass, as shown in
Fig.~5.  This part of our study is completed by the results for
the constituent quark masses, the Polyakov loop and the charge
chemical potential, see Figs.~\ref{FIG:nn} and~\ref{FIG:mue}.

We have computed the masses of the pseudo-Goldstone modes and of
the $\sigma$-mode. The behavior of the masses as a function of the
temperature for different values of the quark chemical potential
are shown in Figs.~\ref{FIG:pp} and~\ref{FIG:sm}. Furthermore we
have investigated on the possibility of existence of bound states
with the quantum numbers of the pions above the chiral critical
temperature. To this end we have compared the computed pion mass
with twice the constituent quark mass. The result is shown in
Fig.~\ref{FIG:kkk}. We have found that for any value of the quark
chemical potential there exists a window in temperature, above the
chiral restoration temperature, in which the pions are lighter
than a state formed by a free quark and a free antiquark. The
aforementioned window shrinks as $\mu$ is increased. Our results
show that even above the critical temperature a bound state with
the pions quantum numbers can be formed. This is summarized in
Fig.~\ref{FIG:ccc}.

In this work we have not studied independently the role of the
chemical potential on the suppression of the pion condensate. This
can be done by leaving the isospin chemical potential as a fixed
and free parameter, and varying only the mean quark chemical
potential. A further natural extension of our work is the study of
the $2+1$ flavor PNJL model with meson condensation, with and
without neutrality conditions implemented. Moreover, it would be
interesting to include a color superconductive state in the
neutral model~\cite{Rapp:1997zu} (studies of the superconductive
state without neutrality conditions can be found in
Refs.~\cite{Roessner:2006xn,Ciminale:2007ei}). Even more one
should compute the transport coefficients of quark matter just
above the chiral transition temperature.  We leave these projects
to future works.


\end{document}